\pdfoutput=1
\documentclass[11pt]{article}
\usepackage{times}
\usepackage{latexsym}
\usepackage[T1]{fontenc}
\usepackage[utf8]{inputenc}
\usepackage{microtype}
\usepackage{inconsolata}
\usepackage{bussproofs}
\usepackage{amsmath}
\usepackage{amssymb, mathrsfs}
\usepackage{tikz}
\usepackage{pgfplots}
\usepackage{subcaption}
\usepackage{tikz-dependency}
\usepackage{hyperref}
\pgfplotsset{compat=1.17}
\usetikzlibrary{positioning}

\newcommand{\lf}{\textsc{lf}}

\newcommand{\xvariable}{{\bf x}}

\newcommand{\zvariable}{{\bf z}}
\newcommand{\cvariable}{{\bf c}}

\newcommand{\pvariable}{{\bf p}}

\newcommand{\gvariable}{{\bf g}}

\newcommand{\wvariable}{{\bf w}}

\newcommand{\condsep}{\ |\ }

\newcommand{\xjack}{\xvariable_{jack}}

\newcommand{\xjill}{\xvariable_{jill}}

\newcommand{\opand}{\textbf{\em and}}
\newcommand{\opor}{\textbf{\em or}}

\title{\bf Logical Discrete Graphical Models Must Supplement Large Language Models for Information Synthesis \\ 
\vspace{25pt}
}
\author{
    {\Large Greg Coppola}
    \\
    {\em coppola.ai} \\
    Research. Develop. Meme.
}
\date{\today}
\begin{document}
\maketitle
\begin{abstract}
\noindent
Given the emergent reasoning abilities of large language models, \emph{information retrieval} is becoming more complex.
Rather than just \emph{retrieve a document}, modern information retrieval systems adverstise that they can \emph{synthesize an answer} based on potentially many different documents, conflicting data sources, and using \emph{reasoning}.
We review recent literature and argue that the \emph{large language model} has crucial flaws that prevent it from on its own ever constituting \emph{general intelligence}, or answering general information synthesis requests.
This review shows that the following are problems for large language models: hallucinations, complex reasoning, planning under uncertainty, and complex calculations.
We outline how logical discrete graphical models can solve all of these problems, and outline a method of training a logical discrete model from unlabeled text.
\end{abstract}

\tableofcontents
\section{Introduction}
Given the emergent reasoning abilities of large language models, \emph{information retrieval} is becoming more complex.
Rather than just \emph{retrieve a document}, modern information retrieval systems adverstise that they can \emph{synthesize an answer} based on potentially many different documents, conflicting data sources, and using \emph{reasoning}.
We review recent literature and argue that the \emph{large language model} has crucial flaws that prevent it from on its own ever constituting \emph{general intelligence}, or answering general information synthesis requests.
This review shows that the following are problems for large language models: hallucinations, complex reasoning, planning under uncertainty, and complex calculations.
We outline how logical discrete graphical models can solve all of these problems, and outline a method of training a logical discrete model from unlabeled text.
\section{Background}
\subsection{Theorem-Proving and Turing Computability}
First-order theorem proving is closely related to the notion of computation itself.
First-order logic is a system of logic in which we can express mathematics, and thus the theory of probability, and thus science.
There are several well-known complete calculus systems for first-order logic \cite{Godel1931,Gentzen1934}.
Any computer program can be represented as a proof in the \emph{typed lambda calculus} of \cite{Church1936}.
Thus, proving a theorem in first-order logic is related to the \emph{halting problem} and thus is semi-decidable, i.e., not decidable \cite{Turing1937, Church1936}.

\subsection{Question-Answering as Theorem-Proving}
\subsubsection{Before Large Language Models}
Before LLM's, a major problem was the inability to parse \emph{open-domain text}.
That is, in order to do symbolic theorem-proving directly, pipelines would require accurate \emph{syntactic parsing} as an earlier stage in the pipeline.
But, before LLM's, it was never possible to make a parser that would work on arbitrary text, often called the \emph{open domain}.
An important project for which the difficulty of open-domain parsing was important was the \cite{lenat1990blk} project, which sought to build a knowledge base of semantic knowledge, but could never annotate enough data.

\subsubsection{The Effect of Large Language Models}
The LLM seemed to side-step this entire problem, by learning world-knowledge in an indirect way, by simply predicting $n$-grams, and learns a knowledge representation as a by-product of training.
Now, since the LLM can effectively, but in an indirect way, do \emph{open-domain parsing}, the situation has changed dramatically.
It is now possible to consider completing the \cite{lenat1990blk} project, which is to discover and extract the knowledge rules that underly human thinking and expression.

\subsubsection{Probabilistic Theorem-Proving}
The thesis of this work is that we should formalize the idea of \emph{answering a question} as the \emph{proving of a theorem} in a logical language extending the \emph{first-order logic} calculus.
\cite{Pereira1987} showed how to handle deterministic theorem-proving in a system of \emph{Horn Clauses}, which we review in Section \ref{s:horn_clauses}.
In \cite{Coppola2024Logical}, we show how to not just prove these theorems but to assign \emph{probabilities} to them.
In \cite{coppola2024categorization} and we show how the logical fragment of Horn clauses fits into increasingly larger fragments, finishing with a \emph{complete} calculus.
\section{A Logical Graphical Model}
In this section, we outline the concept of a \emph{logical graphical model} as described in \cite{Coppola2024Logical}, and analyzed further in \cite{coppola2024categorization}.

\subsection{First-Order Logical Fragments}
We begin from the premises that the first-order logic formalization is sufficient for understanding mathematical and scientific reasoning, as is typically assumed in the philosophy of science, e.g., \cite{Pelletier2000}.
\subsubsection{Horn Clauses}
\label{s:horn_clauses}
One very important subclass of all first-order logic is the system of \emph{Horn Clauses} of the form:
\begin{equation}
    \forall x_1, \ldots, \forall x_k, \Bigg[ \bigwedge_{i=1}^{n} P_i(x_1, \ldots, x_k) \Bigg] \rightarrow C(x_1, \ldots, x_k)
    \label{e:basic_horn}
\end{equation}
A database of sentences of this form can be compiled into a smaller number of more complicated setnences of the form:
\begin{equation}
    \forall x_1, \ldots, \forall x_k, \Bigg[ \bigvee_{i=1}^{n}  \bigwedge_{i=1}^{n} P_i(x_1, \ldots, x_k) \Bigg] \rightarrow \Bigg[ \bigwedge_{i=1}^{n}  C(x_1, \ldots, x_k) \Bigg]
    \label{e:final_horn}
\end{equation}
That is, the premise is in \emph{disjunctive normal form}, and the conclusion is a \emph{conjunction} of basic terms.
The system of Horn Clauses we called the \emph{Direct} fragment.
This is not just the fastest of the fragments, but actually a very fast fragment in absolute terms.
That is, if we obey the \emph{safety} restriction on quantified variables from \emph{Datalog} \cite{abiteboul1995foundations,Pereira1987}, then the complexity of proving a theorem is linear in the number of variables in the graph \cite{abiteboul1995foundations}.
Since the complexity of inference cannot be \emph{less} than linear in the complexity of the graph itself, we say this fragment is \emph{fast}.

\subsubsection{The Query Fragment}
The so-called \emph{Query} fragment allows us to refer to \emph{existentially} bound variables in the premise that do not occur in the conclusion:
\begin{equation}
    \forall x_1, \ldots, \forall x_k, \Bigg[ \exists x_{k+1}, ..., x_K, \bigwedge_{i=1}^{n} P_i(x_1, \ldots, x_k, x_{k+1}, ..., x_K) \Bigg] \rightarrow C(x_1, \ldots, x_k)
\end{equation}
This allows us to express, for example, that if $x_1$ and $x_2$ both \emph{want} the same exclusively held thing $x_3$, then this means $x_1$ and $x_2$ are in \emph{competition}:
\begin{equation}
    \forall x_1, x_2, \Big[ \exists x_3 want(x_1, x_3, \wedge want(x_2, x_3)) \Big] \rightarrow competing(x_1, x_2)
\end{equation}
This amounts to \emph{existentially querying} to see if any object $x_3$ can be found at all.
If we make the simplifying assumption that this query only should return true if $want(x_1, x_3)$ and $want(x_2, x_3)$ are in a concrete database that we can query in time only depending on the database itself, this ability to existentially quantify does not add significantly to the run time compared to the direct model.
However, if we want to be able to search all proofs that can recursively prove the premise, this can lead to an exponential blow-up in complexity, as discussed in \cite{coppola2024categorization}.
Ultimately, the ability to existentially quantify over variables in the premise is useful for expressing knowledge patterns we are interested in, and the speed of the fragment can be fast if we do not seek recursive proofs.

\subsubsection{The Planning Fragment}
The final decidable fragment considered in \cite{coppola2024categorization} is the \emph{Planning} fragment. This allows statements with a disjunction in the conclusion:
\begin{equation}
    \forall x_1, \ldots, \forall x_k, \Bigg[ \bigvee_{i=1}^{n} \bigwedge_{i=1}^{n} P_i(x_1, \ldots, x_k) \Bigg] \rightarrow \bigvee_{i=1}^{n} C(x_1, \ldots, x_k)
    \label{s:planning_form}
\end{equation}
In \cite{coppola2024categorization}, we discuss in detail how this relates to \emph{planning under uncertainty}.
In order to be able to \emph{discharge} or \emph{eliminate} the \emph{or} symbol $\vee$, we need to use the rule that \cite{Prawitz1965} calls \emph{$\vee$-Elimination}.
The ability to \emph{reason by cases} amounts to to boolean satisfiability, or tautology, and thus is \emph{NP-hard} \cite{Cook1971}.
So, while the \emph{Direct} fragment is solvable in \emph{linear}, the \emph{Planning} fragment is instead \emph{NP-hard}, which is believed to be $\Omega(2^N)$ in the number of variables $N$.

\subsubsection{The Full Undecidable Calculus}
\cite{Prawitz1965} presents a calculus with exactly \emph{twelve} deduction rules, corresponding to one \emph{introduction} and one \emph{elimination} for each of the six logical connectives $\vee$, $\wedge$, $\forall$, $\exists$, $\rightarrow$ and $\bot$.
By bounding the run-time of the other fragments, we show that the \emph{undecidability} of first-order logic must come from the deduction rules that \cite{Prawitz1965} calls \emph{improper}, which we instead propose to call \emph{complex} for the modern linguistic context.
This is in other words a mathematical explanation of \cite{Kahneman2011ThinkingFast}'s distinction between \emph{thinking} that is \emph{fast} versus \emph{slow}.
Also, we show why it would seem that it is \emph{not} sensible to try to fit \emph{all} first-order reasoning into the paradigm of a single pass of inference in a graphical model.
That is, the reason is that in the more \emph{complex} deduction rules, there is a change in assumptions (or other hard to manage change) that means that the \emph{assumptions} must be changed.
But, in a single pass of inference in a graphical probabilistic model, each variable must either take one value or the other.
Thus, while there are more inference rules that can be implemented than just those in \cite{Coppola2024Logical}, we conclude that not \emph{all} first-order deduction rules can be implemented in the context of one pass of inference in a graphical model.
Instead, one must use \emph{book-keeping} to keep track of which \emph{Direct} and \emph{Query} inferences can be derived for each change of assumptions.
But, we propose that we can use the same book-keeping mechanism that \cite{Prawitz1965} did for deterministic theorem-proving.

\subsection{Three Levels of Graphical Structure}
\subsubsection{Knoweldge Graph}
\paragraph{Overview}
Recently, an important research direction has been the use of \emph{retrieval augmented generation} \cite{lewis2013combined:2020} to the creation of \emph{knowledge graphs}.
That is, if an LLM can be trained to read a document, and extract the information for the user from the document, and return a summary of that.
The observation then is that if we have already retrieved and parsed a document, we can simply store the answer (assuming a baseline level of syntactic analysis ability) in a structured database.
A knowledge graph can be viewed as a set of \emph{tables}, making up a \emph{database}.
Each table corresponds to a verb, e.g., $likes$.
If the verb $likes$ has the roles \textsc{subj} and \textsc{dobj}, then there would be two columns in the table, and one entry per pair of people $x_1$ and $x_2$, such that $likes(x_1, x_2)$.
\paragraph{Problem}
In the interpretation, a central feature is that a row is discretely either \emph{in} the database or \emph{out} of it.
All statements about future beliefs must definitely be \emph{probabilistic}, assuming the future has not ``happened'' yet.
But, also beliefs about the present and past should be treated as probabilistic from a scientific perspective, although from the user's perspective certain facts can be considered true with essentially probability $1$, e.g., that \emph{New York City} is in \emph{New York State}.
But, if we ask a historical question, like \emph{did Julius Caesar really escape a pirate ship?}, or \emph{is the 10 Commandments a historical story?}, we cannot say with probability $1$ whether these sentences hold.
Similarly, questions like \emph{is product X in stock at store Y} and \emph{will it rain tomorrow?} cannot generally be given deterministic answers.
This corresponds to saying that the \emph{knowledge graph} must not be simply a discrete database entry, in which an entity is either a member or not, but must be a \emph{probabilistic} database, in which we can ask about the \emph{probability} of a semantic relationship holding.

\subsubsection{Implication Graph}
Chomsky was famously fond of quoting Humboldt's aphorism that language makes {\em infinite use of finite means} \cite{Chomsky1965Aspects}.
The {\em implication graph} allows us to estimate probabilities for an {\em unbounded} number of {\em propositions} $\pvariable$ based on finite parameters $\Psi$, by defining weights over {\em predicate patterns}, rather than relationships between {\em concrete entities}.
That is, we learn a general link between $\xjack$ {\em liking} $\xjill$ and $\xjack$ {\em dating} $\xjill$, and this can apply to $\cvariable_{jack1}$ or $\cvariable_{jack2}$ or $\cvariable_{jill1}$ or $\cvariable_{jill2}$, etc., and so make {\em infinite use} of {\em finite means}.
The logical and statistical \emph{implications} in the system are expressed as \emph{Horn Clause} sentences of the form of \ref{e:final_horn}.

\subsubsection{Proposition Graph}
A \emph{proposition graph} is an instantiation of the \emph{implication graph} to answer a particular question.  
For an unbounded set of {\em entities}, there are an unbounded number of {\em possible propositions} $\pvariable$, many of which will never be relevant.
For example, consider the predicate of {\em is President of the United States}.
This only applies in practice to one person, but could, in principle, apply to billions.
Thus, storing all {\em possible} propositions in hard disk memory would not be possible.
Thus, the proposition graph is lazily constructed at answer-time.
We have two options in the system for estimating the probability of a proposition $\pvariable$.
The first is that the probability for the proposition is {\em alredy computed} before the {\em query} is issued.
In the example of {\em $\xvariable_{person}$ is the President of the United States}, this can be set to {\em true} in long-term storage for the unique individual who occupies this slot.
For anyone {\em else}, we can use {\em generic} reasoning, like {\em there is only one President}, and {\em right now the President is someone else}, etc., to answer {\em no}.

\subsection{Logical Boolean Algebra}
A central feature of the paradigm we are proposing, in which we follow the rules of logic \cite{Frege1879, Godel1931}, especially that of the \emph{natural deduction calculus} \cite{Gentzen1934, Prawitz1965}.
In all of these logics, the two primary boolean connectives are \emph{and} $\wedge$ and \emph{or} $\vee$, and these form a \emph{boolean algebgra} \cite{boole1854investigation}.
Because the factor types $\Psi_\opand$ and $\Psi_\opor$ always alternate, we have a {\em bipartite graph}.
Suppose $\pvariable_1, ..., \pvariable_n$ are each {\em propositions}.
Then we say 
\begin{equation} \gvariable = \left\{\pvariable_1 \wedge ... \wedge \pvariable_n\right\} \end{equation}
is a {\em proposition group}, which are interpreted as {\em conjoined}.
Then, the two types of variables in the graph then are:
\begin{enumerate}
    \item $\pvariable$, which represents a {\em single proposition}
    \item $\gvariable$, which represents a {\em conjoined proposition group}
\end{enumerate}
In \cite{Coppola2024Logical}, for many purposes in the graphical model (e.g., {\em message passing} calculuations), we can abstract over whether a node is $\gvariable$ and $\pvariable$, and we refer to {\em generic graphical nodes} as $\zvariable$, which greatly simplifies visualization and implementation.

\subsection{Conjunction Nodes}
\label{seq:conjunction-nodes}
\subsubsection{Deterministic Definition}
The {\em conjunctive} factor $\Psi_\opand$ is defined in terms of the {\em \opand\ gate}:
\begin{equation} \opand(\pvariable_1, ..., \pvariable_n) = \pvariable_1 \wedge ... \wedge \pvariable_n \end{equation}
Then:
\begin{equation}
    \Psi_{\opand}(\gvariable \condsep \pvariable_1, \ldots, \pvariable_n) = 
    \begin{cases} 
        1 & \text{if } \gvariable == \opand(\pvariable_1, \ldots, \pvariable_n), \\
        0 & \text{otherwise}
    \end{cases}
    \label{eq:op_and_determinstic}
\end{equation}
This is used in the completeness proof, but also in learned models.

\subsubsection{Conjunction and Higher-Order Feature Representations}
Let us meditate on the fact that the $\Psi_\opand$ factor is {\em always} deterministic, i.e., we do not train this even when we are interested in statistical inference.
One way to justify this is that, intuitively, \opand's role is to create {\em higher-level features}, between which we can learn relationships.
This is like a {\em discrete} analog to the {\em higher-level} features that {\em multi-layer networks} learn \cite{Rumelhart1986, LeCun1989}.
For example, suppose we are given the information about $like(\xjack, \xjill)$ and $like(\xjill, \xjack)$ as {\em simple} features to predict $date(\xjack, \xjill)$.
Supposing the relevant higher level feature is $like(\xjack, \xjill) \land like(\xjill, \xjack)$, one option is run these features through a multi-layer network, which will be able to {\em learn} this feature, in a {\em differentiable} way.
But, this higher-level feature emerges as an effectively {\em emergent} behavior.
This leads to the problem of {\em interpretability} \cite{Hinton2015, Ribeiro2016, Lundberg2017}.
The {\em logical graphical model} is another interpretation of {\em interpretability}, because the features must be {\em explicitly} conjoined in order to work.
The problem is not one of {\em interpreting} the model, but {\em constructing} the model in the first place, since the individual {\em function names} and {\em role labels} underlying logical ``language'' are latent \cite{Steedman1996}, and presumably would be discovered through something analogous to {\em category splitting} in a generative model \cite{Petrov2006}.

\subsection{Disjunction Nodes}
\subsubsection{Deterministic Logical Reasoning}
The deterministic {\em disjunctive} factor $\Psi_{\opor}$ used for the {\em completeness proof} (also see \cite{Coppola2024}), is defined in terms of the {\em \opor\ gate}:
\begin{equation} 
    \opor(\gvariable_1, \ldots, \gvariable_n) = \gvariable_1 \vee \ldots \vee \gvariable_n 
\end{equation}
The {\em deterministic} version of \opor, used in the {\em completeness} proof, and can be used any time we want exact logical \opor, is defined as:
\begin{equation}
    \Psi_{\opor}(\pvariable \condsep \gvariable_1, \ldots, \gvariable_n) = 
    \begin{cases} 
        1 & \text{if } \pvariable == \opor(\gvariable_1, \ldots, \gvariable_n), \\
        0 & \text{otherwise}
    \end{cases}
    \label{eq:op_or_determinstic}
\end{equation}
When interested in {\em statistical inference}, we {\em learn} this model, as discussed in Section \ref{sec:learned_model}.

\subsubsection{Probabilistic Reasoning}
\label{sec:learned_model}
\paragraph{Model Structure}
For the {\em learned model}, we model $\Psi_\opor$ using {\em linear exponential} model.
For a boolean variable \( \pvariable \) with boolean features \( \gvariable_1, ..., \gvariable_n \), the {\em factor potential} has the form:
\begin{equation} 
\Psi_{\opor} (\pvariable \condsep \gvariable_1, ..., \gvariable_n) = \exp{ \left\{\sum_{i=1}^{n} {\wvariable \cdot \phi(\pvariable, \gvariable_i)}\right\}}
\label{e_linear_exponential}
\end{equation}
Here, \( \wvariable \) is a weight vector, and \( \phi(\pvariable, \gvariable_i) \) is a feature function discussed in \cite{Coppola2024Logical}.
The probability \( P(\pvariable \condsep \gvariable_1, ..., \gvariable_n) \) is obtained by normalization over the two possible values for \( \pvariable \in \left\{0, 1\right\} \):
\begin{equation} 
P(\pvariable = p \condsep \gvariable_1, ..., \gvariable_n) = \frac{\Psi_{\opor} (p \condsep \gvariable_1, ..., \gvariable_n)}{\Psi_{\opor} (1 \condsep \gvariable_1, ..., \gvariable_n) + \Psi_{\opor} (0 \condsep \gvariable_1, ..., \gvariable_n)} 
\end{equation}

\paragraph{Similarity Between Linear Exponential and Disjunction}
To underline the similarity between the {\em linear exponential model} and {\em disjunction} $\Psi_\opor$, consider how we implement $\opor$ using a log-linear model.
That is, the dependence of $Y$ on $X_1$ and $X_2$ can be expressed as:
\begin{equation}
    P(Y=1|X_1, X_2) = \frac{1}{1 + \exp(-(\beta_0 + \beta_1X_1 + \beta_2X_2))}
\end{equation}
If we set $\beta_0 = -0.5$, a negative bias, and $\beta_1 = \beta_2 = 1$, then this predicts $Y = 1$ if either $X_1 = 1$ or $X_2 = 1$, but $Y=0$ otherwise.
That is, it implements \opor, and this technique scales for $n>2$.
\section{Hallucinations}

\subsection{The Problem of Hallucination}
While the {\em large language model} is widly popular for its ability to learn complex kinds of {\em knowledge} and even some {\em reasoning} from {\em unlabeled text},
the primary empirical user complaint with {\em large language models} is that of {\em hallucinations} \cite{SutskeverHuang2023}.
That is, a {\em large language model} can return answers that are not ``supported by the training set'' when judged by a {\em human evaluator}.
This lack of reliability greatly limits throughput, because it requires all output of a {\em large language model} to be double-checked by the user.
\subsection{Relation to Causality}
Our analysis is that the concept of \emph{hallucination} is closely related to the concept of \emph{causality}.
In other words, if we have a model that can \emph{explain why} it is giving an answer, assuming that this method of explanation is sensible, can never hallucinate, because it can always give the basis for its beliefs, and, we are asuming, the explanation is sensible.
Thus, a model which is both sensible and aware of causality cannot hallucinate.

\subsection{Compared to Existing Solutions}
\subsubsection{Discriminative Fine-Tuning}
The earliest used solution for controlling LLM behavior was discriminative \emph{fine-tuning}, in which a discriminative model is used on top of the generative pre-trained model to predict a certain task \cite{radford2019language, radford2019language}.
On its own this did not completely prevent hallucinations.
Compared to the paradigm of generative pre-training and discriminative fine-tuning, the \emph{logical graphical model} approach is notable for the fact that it is a \emph{generative} model, but it \emph{does not hallucinate} by construction, because it explains causality.
Thus, the logical graphical model is a generative model that can avoid hallucinating, without even needing a discriminative stage at all.

\subsubsection{Retrieval-Augmented Generation}
Another important technique is \emph{retrieval-augmented generation}, in which a discrete document is used to answer the question \cite{lewis2013combined:2020}.
This is a useful technique, but has the draw back that the individual document used must be treated as \emph{ground truth}.
There is no way with basic \emph{retrieval augmented generation} to store contradictory information, and synthesize it in a coherent way.
\section{Complex Reasoning}
\subsection{Exact Reasoning}
\cite{schick2023toolformer} showed that it is more accurate to use an LLM to realize that it needs to use a deterministic calculator, than it is to simply let the LLM try to compute each number as a probabilistic prediction.
Indeed, it stands to reason that this would be not only more accurate but also cheaper.
In the case of the logical graphical model, because we are converting to a discrete space, we can easily include function calling in a number of ways.
Indeed, according to theoretical computer science, the computation of an exact function like a calculator can be expressed as a logical graph \cite{Church1936}.
\subsection{Long ``Chains'' of Reasoning}
A popular strain of reasoning lately began with \emph{chain-of-thought} \cite{wei2022chain}, then \emph{tree of thought} \cite{yao2023tree}, the \emph{chain-of-abstraction} \cite{gao2024efficient}.
In each case, the use of deterministic, hard-coded \emph{prompt modules} is used to either prompt another LLM response based on the previous one, or else to take over the computation entirely.
\emph{Chain-of-Abstraction}, for example, says they train LLMs to first decode reasoning chains with abstract placeholders, and then call domain tools to reify each reasoning chain by filling in specific knowledge.
In other words, the LLM is effectively being used as a \emph{syntactic parser}, after which computation is fully symbolic.
The problem with the \emph{Chain of X} paradigm, is that it does \emph{not} say what the relevant \emph{abstraction} is.
In our case, we propose that the key data structures and abstractions with which to think about computation are: first-order logic, discrete graphical models and boolean algebra.

\subsection{Multiple Points of View}
\cite{Hinton2023} has suggested that large language models are unable to understand that a user may want to see the same issue analyzed from \emph{multiple points of view}.
\cite{Steedman2022} has proposed that these models are not exactly doing logical reasoning.
Indeed, the mathematical task of a large language model is to predict a \emph{sequence} of text is to predict a continuation passage.
And, it is well-known that a large language model can continue \emph{in the voice} of arbitrary chracters.
But, this is only a surface-level appearance of understanding.
When compared with the \emph{logical graphical model}, which can truly reason based on different assumptions, we see that the shallow imitation of true multi-viewpoint thinking of LLM's will not allow us to learn truly deep, complex patterns about the world.
That is, because the logical graphical model is discrete, and so individual assumptions can easily be set to \emph{true} or \emph{false}.
\section{Estimation and Inference}
\label{s:estimation}
In this section, we discuss how to estimation and inference with a graphical model.

\subsection{Estimation}
\paragraph{LLM's}
The key fact about the LLM is that the LLM is able to \emph{compress} the dataset \cite{SutskeverObservation}.
What makes the original LLM tractable to estimate is that it just predicts the next sentence $\xvariable_{n+1}$ based on the previous sentence $\xvariable_n$, and the parameters $\phi$:
\begin{equation}
    P(\xvariable_{n+1} \condsep \xvariable_{n}, \phi)
\end{equation}
This is straightforward to optimize because all variables $\xvariable_n$ are \emph{full observable}.
\paragraph{Generating Parse Structure}
In order to replicate this generative with a logical model, we will need to be able to generate a form that has a logical interpretation.
In order to generate the sentence, we can opt to use the strategy of creating a \emph{labeled dependency parse} that is assumed to have generated the sentence.
We will represent a parse as $y$ and write $\lf(y)$ to indicate the possibly complex sentence in the first-order language that $y$ is semantically analyzed as.
That is, $\lf(y)$ determinisitically depends on $y$.
This dependency parse is \emph{latent}, in the sense that it is not observed, unlike the surface form of the sentence.
With a logical model, we separate this into two parts as:
\begin{equation}
    P(\xvariable_{n+1} \condsep \xvariable_{n}, \phi, \theta) \propto
    \Sigma_{y \in C(\xvariable_{n+1})} P(y, \xvariable_{n+1} \condsep \xvariable_{n}, \phi) \cdot P(\lf(y) \condsep \theta )
\end{equation}
Here, $C(\xvariable_{n+1})$ is the set of candidate parses for the sentence $\xvariable_{n+1}$, and $\theta$ is a set of logical parameters that is used to score the prior likelihood that 
\cite{Coppola2024Logical} shows how we can do inference on $P(\lf(y) \condsep \theta)$ using \emph{loopy belief propagation} \cite{pearl1988probabilistic, murphy1999loopy}.
This is not guaranteed to be true, but we find that loopy belief propagation converges well in practice.

\paragraph{Expectation Maximization}
Because the parses are \emph{latent}, they cannot be observed directly like the surface form of a sequence of tokens that the LLM trains on.
Thus, we must estimate the parses with some variant of \emph{expectation maximization} \cite{dempster1977maximum}.
While a ``full'' implementation of expectation maximization can be very complicated, there are very simple variants of the algorithm, including \emph{$n$-best expectation maximization}, where only a finite number of options from a first-stage syntactic parser, are rescored according to the language model \cite{charniak-johnson-2005-coarse}.
One can even do $1$-best EM, and simply take the most likely parse, but with the logical model also having input about the likelihood of the interpretation.

\subsection{Inference}
Assuming that we have a labeled dependency parser, when given a query, we can simply \emph{parse} that query to a logical representation of a question.
\begin{equation}
    \arg\max_{y \in C(\xvariable_{n+1})} P(y \condsep \xvariable_{n+1}, \xvariable_{n})
\end{equation}
Then, the \emph{logical form} implied by the syntactic analysis $y$, called $\lf(y)$, will contain a mixture of \emph{assumptions}, which we denote $A_y$, and \emph{questions}, which we denote $Q_y$.
Then, the system can assume $A_y$, and answer the questions $Q_y$ in logical space.
Suppose the answer is $b$. Then, we can generate a \emph{surface form} of $b$, which we can call $\textsc{sf}(b)$, generated according to a language model, conditioned on the logical form $b$, using the classic concepts of dialog theory \cite{allen1995planning, litman1986plan}.
\bibliographystyle{apalike}
\bibliography{bibtex}
\end{document}